\documentclass[10pt,twocolumn,prl,aps,amssymb,amsmath,tightenlines,showpacs]{revtex4}

\newcommand{\cC}{\ensuremath{\mathcal{C}}}
\newcommand{\cP}{\ensuremath{\mathcal{P}}}
\newcommand{\cT}{\ensuremath{\mathcal{T}}}

\begin{document}

\title{$\cP\cT$-symmetric quantum state discrimination}

\author{Carl~M.~Bender${}^1$, Dorje~C.~Brody${}^2$,
Jo\~ao Caldeira${}^3$, and Bernhard~K.~Meister${}^4$}

\affiliation{${}^1$Physics Department, Washington University, St.
Louis, MO 63130, USA\\
${}^2$Mathematics Department, Imperial College, London SW7 2AZ, UK\\
${}^3$Blackett Laboratory, Imperial College, London SW7 2AZ, UK\\
${}^4$Department of Physics, Renmin University of China, Beijing 100872, China}

\date{\today}

\begin{abstract}
Suppose that a system is known to be in one of two quantum states, $|\psi_1
\rangle$ or $|\psi_2\rangle$. If these states are not orthogonal, then in
conventional quantum mechanics it is impossible with one measurement to
determine with certainty which state the system is in. However, because a
non-Hermitian $\cP\cT$-symmetric Hamiltonian determines the inner product that
is appropriate for the Hilbert space of physical states, it is always possible
to choose this inner product so that the two states $|\psi_1\rangle$ and $|
\psi_2\rangle$ are orthogonal. Thus, quantum state discrimination can, in
principle, be achieved with a single measurement.
\end{abstract}
\pacs{11.30.Er, 03.65.Ca, 03.65.Xp}
\maketitle

The problem of quantum state discrimination is important in many applications of
quantum information technology. Typically, one wants to extract information that
is encoded in the unknown state of a quantum system. Therefore, one measures an
observable, the outcome of which provides some information about the state of
the system. Solving this problem amounts to finding (i) the optimal choice for
the observable, and (ii) the optimal strategy to infer the state of the system,
given the outcome of the measurement. In this paper we discuss the following
idealized binary state-discrimination problem: An experimentalist who wishes to
determine the state of the system is given the {\em a priori} information that
the system is in one of two possible states, $|\psi_1\rangle$ or $|\psi_2
\rangle$, which are not orthogonal. It is not possible to ascertain with
certainty the state of the system with a single measurement. However, repeated
measurements on a single system are not in general permissible because a
measurement can change the state of the system. Thus, to identify the state of
the system with a high confidence level a large number of identically prepared
samples may be needed.

There is an extensive literature on various approaches to quantum state
discrimination; see, for example, the recent review articles \cite{R1,R2} and
references cited therein. If the experimentalist knows that the state of the
system is either $|\psi_1\rangle$ or $|\psi_2\rangle$, then the task reduces to
making a quantum binary decision. Depending on the criteria of optimality,
various optimal solutions to the detection problem have been found. The result
of Helstrom \cite{R3}, in particular, provides a bound on the error probability
associated with a single measurement. If the two possible input states are close
to each other so that $|\langle\psi_1|\psi_2\rangle|^2 \approx 1-\epsilon$ with
$\epsilon\ll1$, then the result of a single measurement is hardly conclusive.
However, if there is a large number of identically prepared samples, all of
which are either in the state $|\psi_1\rangle$ or $|\psi_2\rangle$, then by
following the optimal strategy of \cite{R4} for sequential measurements, one can
asymptotically discriminate the state with a high confidence level. On the other
hand, preparing a large number of identical samples can be costly, and
determining the state will be time consuming.

This problem has a simple classical analog: One is told that a given coin is
either (i) a fair coin and that the probability of heads is exactly 50\%, or
else that (ii) the coin is unfair and that the probability of heads is 50.1\%.
To distinguish experimentally between these two possibilities requires a lengthy
string of coin tosses before one can make a decision with confidence, and one
can never be absolutely certain which possibility is the truth.

The purpose of the present paper is to propose an alternative approach to
quantum state discrimination that allows one to determine the state with
certainty by just a single measurement. The idea is to introduce a carefully
chosen complex $\cP\cT$-symmetric (space-time reflection symmetric) Hamiltonian.
If the $\cP\cT$ symmetry of such a Hamiltonian is not broken, then the
eigenvalues are real. Such a Hamiltonian $H$ determines the inner product in the
Hilbert space on which it is defined, and relative to this inner product its
eigenvectors are orthogonal \cite{R5,R6,R7}. If $H$ is chosen correctly, then
the inner product of the two states $|\psi_1\rangle$ and $|\psi_2\rangle$ is
arbitrarily small, and there exists an observable that perceives these two
states as being orthogonal. In fact, such an observable is unitarily equivalent
to $H$ itself.

In this paper we propose two different but related solutions to the binary
state-discrimination problem. The first is simply to apply a binary measurement
in a complex direction. Depending on the outcome, the state of the system can
then, in principle, be determined with certainty. Of course, the practical 
implementation of a complex measurement can be challenging. However,
implementing a dynamical evolution governed by a complex Hamiltonian having real
eigenvalues is more amenable experimentally \cite{R8,R9,R10}, so we present a
second solution. Such an evolution can be achieved in a non-Hermitian system in
which there is a delicate and precise balance of loss and gain. This suggests a
second and alternative solution whereby a unitary evolution using a complex
Hamiltonian in a suitably defined Hilbert space is applied so that the two input
states $|\psi_1\rangle$ and $|\psi_2\rangle$ evolve into a pair of states that
are perceived as being orthogonal in the conventional Hermitian inner product
Hilbert space. A real binary measurement can then be applied to distinguish the
states with certainty.

\smallskip\noindent
{\em Solution 1: Finding a $\cP\cT$-symmetric Hamiltonian whose inner product
interprets $|\psi_1\rangle$ and $|\psi_2\rangle$ as being orthogonal}. We
consider the two-dimensional subspace spanned by the two vectors $|\psi_1
\rangle$ and $|\psi_2\rangle$. Let the angular distance between the two states
in the Bloch sphere be $2\epsilon$. Without loss of generality we can
reparametrize the Bloch sphere so that both states lie on the same meridian;
that is, $|\psi_1\rangle$ lies at the angles $(\theta,\phi)$ and $|\psi_2
\rangle$ lies at $(\theta+2\epsilon,\phi)$:
\begin{equation}
|\psi_1\rangle=\left(\begin{array}{c}\cos\frac{\theta}{2}\cr e^{i\phi}\sin
\frac{\theta}{2}\end{array}\right),\qquad
|\psi_2\rangle=\left(\begin{array}{c}\cos\frac{\theta+2\epsilon}{2}\cr
e^{i\phi}\sin\frac{\theta+2\epsilon}{2}\end{array}\right).
\label{e1}\
\end{equation}
We still have the freedom to choose specific values for $\theta$ and $\phi$,
and for simplicity we choose $\phi=-\frac{\pi}{2}$ and $\theta=\frac{\pi}{2}-
\epsilon$.

Let us consider the general $2\times2$ $\cP\cT$-symmetric Hamiltonian \cite{R6}
\begin{eqnarray}
H&=&\left(\begin{array}{cc} re^{i\beta} & s\cr s&r^{-i\beta}\end{array}
\right)\nonumber\\
&=&r\cos\beta\,{\bf 1}+\boldsymbol{\sigma}\cdot\left(s,\,0,\,ir\sin\beta\right),
\label{eq2}
\end{eqnarray}
where the parameters $r$, $s$, and $\beta$ are real and $\boldsymbol{\sigma}$ 
are the Pauli matrices
$$\sigma_1=\left(\begin{array}{cc} 0&1\\ 1&0\end{array}\right),\quad
\sigma_2=\left(\begin{array}{cc} 0&-i\\ i&0\end{array}\right),\quad
\sigma_3=\left(\begin{array}{cc} 1&0\\ 0&-1\end{array}\right).$$ 
This Hamiltonian commutes with $\cP\cT$, where the parity reflection operator
is given by
\begin{eqnarray}
\cP=\sigma_1
\label{eq3}
\end{eqnarray}
and the time-reversal operator $\cT$ is complex conjugation.

For $H$ in (\ref{eq2}) the parametric region of unbroken $\cP\cT$ symmetry in
which the eigenvalues are real is $s^2>r^2\sin^2\beta$. In this region we can 
calculate the $\cC$ operator:
\begin{eqnarray}
\cC=\frac{1}{\cos\alpha}\left(\begin{array}{cc} i\sin\alpha & 1\cr 1 & -i\sin
\alpha\end{array}\right),
\label{eq4}
\end{eqnarray}
where $\sin\alpha=\frac{r}{s}\sin\beta$. Then, using the $\cC\cP\cT$ operator,
we can calculate the bra vectors corresponding to ket vectors. Specifically, we
find that for $|\psi_1\rangle$ in (\ref{e1}) the corresponding $\langle\psi_1|$
is the row vector
\begin{eqnarray}
\langle\psi_1|&=&\frac{1}{\cos\alpha}\left(\cos\frac{\pi-2\epsilon}{4}-\sin
\alpha\,\sin\frac{\pi-2\epsilon}{4},\right.\nonumber\\
&&\left.\qquad-i\sin\alpha\,\cos\frac{\pi-2\epsilon}{4}+i\sin\frac{\pi-2
\epsilon}{4}\right).
\label{eq5}
\end{eqnarray}
Thus, we can calculate the inner product $\langle\psi_1|\psi_2\rangle$, and if
we require that this inner product vanish, we obtain the condition
\begin{eqnarray}
\sin\alpha=\cos\epsilon.
\label{eq6}
\end{eqnarray}

Finally, to distinguish between the two states $|\psi_1\rangle$ and $|\psi_2
\rangle$, we need only construct projection operators that leave one state
invariant and annihilate the other state. To do so we must normalize these
states. A straightforward calculation gives
\begin{eqnarray}
\langle\psi_1|\psi_1\rangle=\langle\psi_2|\psi_2\rangle=\sin\epsilon.
\label{eq7}
\end{eqnarray}
Hence, the {\em normalized} state $|\psi_1\rangle$ is given by
\begin{eqnarray}
|\psi_1\rangle&=&\frac{1}{\sqrt{\sin\epsilon}}
\left(\begin{array}{c} \cos\frac{\pi-2\epsilon}{4}\cr
-i\sin\frac{\pi-2\epsilon}{4}\end{array}\right),\nonumber\\
\langle\psi_1|&=&\frac{1}{\sin^{3/2}\epsilon}\left(\cos\frac{\pi-2\epsilon}{4}
-\cos\epsilon\,\sin\frac{\pi-2\epsilon}{4},\right.\nonumber\\
&&\left.\qquad-i\cos\epsilon\,\cos\frac{\pi-2\epsilon}{4}+
i\sin\frac{\pi-2\epsilon}{4}\right).
\label{eq8}
\end{eqnarray}
The results for $|\psi_2\rangle$ and $\langle\psi_2|$ are obtained by replacing
$\pi-2\epsilon$ with $\pi+2\epsilon$.

We then construct the projection operators
\begin{eqnarray}
|\psi_1\rangle\langle\psi_1|&=&\frac{1}{2\sin\epsilon}\left(
\begin{array}{cc} 1+\sin\epsilon & -i\cos\epsilon \cr -i\cos\epsilon & 
-1+\sin\epsilon\end{array}\right),\nonumber\\
|\psi_2\rangle\langle\psi_2|&=&\frac{1}{2\sin\epsilon}\left(
\begin{array}{cc} -1+\sin\epsilon & i\cos\epsilon \cr i\cos\epsilon & 
1+\sin\epsilon\end{array}\right).
\label{eq9}
\end{eqnarray}
It is straightforward to verify that these operators are $\cP\cT$ {\em
observables} because they are $\cC\cP\cT$-selfadjoint \cite{R5,R6}; that is,
they commute with the $\cC\cP \cT$ operator. Furthermore, these projection
operators constitute a resolution of the identity:
\begin{equation}
|\psi_1\rangle\langle\psi_1|+|\psi_2\rangle\langle\psi_2|={\bf 1}.
\label{eqx}
\end{equation}

The projection operators in (\ref{eq9}) can be expressed as a linear combination
of Pauli sigma matrices,
\begin{eqnarray}
|\psi_1\rangle\langle\psi_1|=\frac{1}{2}{\bf 1}+\boldsymbol{\sigma}\cdot\left(
-\frac{i}{2}{\rm cot}\,\epsilon,\,\,0,\,\,\frac{1}{\sin\epsilon}\right),
\label{eq10}
\end{eqnarray}
and so can the Hamiltonian:
\begin{eqnarray}
H=\sqrt{r^2-s^2\cos^2\epsilon}\,{\bf 1}+\boldsymbol{\sigma}\cdot\left(s,\,\,0,\,
\,is\cos\epsilon\right).
\label{eq11}
\end{eqnarray}
Thus, we see that these operators are equivalent to applying a magnetic field in
a complex direction. A single application of one of the projection measurements
in (\ref{eq9}) distinguishes the states $|\psi_1\rangle$ and $|\psi_2\rangle$
with certainty.

\smallskip\noindent
{\em Solution 2: Finding a $\cP\cT$-symmetric Hamiltonian under which the states
$|\psi_1\rangle$ and $|\psi_2\rangle$ evolve into orthogonal states}. Recent
experimental results in Refs.~\cite{R8,R9,R10} indicate that it may be easier to
implement a non-Hermitian Hamiltonian than to implement a non-Hermitian
observable. In such cases there is an alternative strategy to accomplish state
discrimination: We construct a Hamiltonian under which the two states $|\psi_1
\rangle$ and $|\psi_2\rangle$ evolve into states that are orthogonal under the
conventional Hermitian inner product. We then proceed to make a measurement
using a conventionally Hermitian observable.

In conventional Hilbert space the standard inner product is based on the
Hermitian adjoint (transpose and complex conjugate). Thus, at time $t$ the inner
product is simply $\langle\psi_1|e^{iH^\dag t}e^{-iHt}|\psi_2\rangle$, where $H$
is given in (\ref{eq2}), $H^\dagger$ denotes the Hermitian adjoint of $H$, and
we have taken $\hbar=1$. We use the standard matrix identity to simplify the
exponential of $H$ in (\ref{eq2}):
\begin{equation}
\exp(i\phi\,{\boldsymbol{\sigma}}\!\cdot\!{\bf n})=\cos\phi\,{\bf
1}+i\sin\phi\,{\boldsymbol{\sigma}}\!\cdot\!{\bf n}.
\label{eq12}
\end{equation}
Using this identity, we obtain the result
\begin{eqnarray}
&& \cos^2\alpha\,e^{iH^\dag t}e^{-iHt}\nonumber\\
\vspace{.3cm}
&& \!\!\!\!\!=\left(\begin{array}{cc}\cos^2(\omega
t-\alpha)+\sin^2(\omega t) & -2i\sin^2(\omega t)\sin\alpha\\ 2i\sin^2(\omega
t)\sin\alpha &\cos^2(\omega t+\alpha)+\sin^2(\omega t)\end{array}\right)
\nonumber
\end{eqnarray}
in which $\omega=\sqrt{s^2-r^2\sin^2\beta}$. (Note that in the Hermitian limit
$\alpha\to0$, this becomes the identity matrix $\bf 1$.)

We thus calculate the inner product at time $t$:
\begin{eqnarray}
\langle\psi_1,t|\psi_2,t\rangle&=&\langle\psi_1|e^{iH^\dag t}e^{-iHt}|\psi_2
\rangle\nonumber\\
&=&\cos\epsilon\left[\cos^2\alpha+2\sin^2(\omega t)\sin^2\alpha\right]
\nonumber\\
&&\quad\quad -2\sin^2(\omega t)\sin\alpha.
\label{eq13}
\end{eqnarray}
This inner product vanishes when
\begin{equation}
\sin^2(\omega t)=\frac{\cos^2\alpha\,\cos\epsilon}{2\sin\alpha-2\sin^2\alpha
\cos\epsilon},
\label{eq14}
\end{equation}
which has a solution for $t$ if $\epsilon\neq0$.

Note that the time needed for this evolution becomes arbitrarily small and
approaches 0 as $\cos\alpha\to0$ (or $\alpha\to\pm\frac{\pi}{2}$). This is an
echo of what was found in the case of the non-Hermitian quantum brachistochrone
\cite{R11,R12,R13,R14}. Among all Hermitian Hamiltonians, the Hamiltonian that
achieves the fastest time evolution from a given initial state to a given final
state still requires a nonvanishing amount of time. However, It was shown
Refs.~\cite{R11,R12,R13,R14} that a non-Hermitian $\cP\cT$-symmetric Hamiltonian
can perform this time evolution is an arbitrarily short time.

{}From (\ref{eq2}) we can see that the limit $\alpha\to\frac{\pi}{2}$
corresponds to an application of a magnetic field in a complex direction and
that the imaginary component of this magnetic field $r\sin\beta$ takes its
highest possible value. There may be practical constraints that make it
difficult to realize such a limit, in which case an experimentalist must wait
some time until (\ref{eq14}) is satisfied. At this point, a Hermitian projection
measurement can be applied to distinguish between the two possible input states.

In summary, we have presented two alternative ways to distinguish between a pair
of nonorthogonal pure quantum states with a single measurement. To do so, we
have exploited the complex degrees of freedom made available by $\cP\cT$
symmetry. If one of these strategies can be implemented, then there are
considerable benefits in the area of quantum information theory. For example, in
quantum computation it is known that an unstructured database search can be
mapped to the problem of distinguishing exponentially close quantum states
\cite{R15}.  The reformulation of the database search can also be achieved using
the method described here to search a database exponentially fast. This is
because the method presented here can be applied to distinguish fast and
accurately any pair of distinct states. It would be of interest to investigate
whether the present scheme can be extended to distinguish a pair of mixed
quantum states.

CMB thanks the U.S.~Department of Energy for financial support.

\end{document}